\documentstyle[12pt,equation]{article}
\begin{document}
\newcommand{\be}{\begin{equation}}
\newcommand{\ben}{\begin{subequations}}
\newcommand{\een}{\end{subequations}}
\newcommand{\beq}{\begin{eqalignno}}
\newcommand{\eeq}{\end{eqalignno}}
\newcommand{\ee}{\end{equation}}
\newcommand{\wt}{\widetilde}
\newcommand{\stw}{\sin^2 \! \theta_W}
\newcommand{\ctw}{\cos^2 \! \theta_W}
\newcommand{\sw}{\sin \! \theta_W}
\newcommand{\cw}{\cos \! \theta_W}
\newcommand{\dmchi}{\mbox{$\Delta m_{\tilde {\chi}}$}}
\newcommand{\mchi}{\mbox{$m_{\tilde {\chi}_1^0}$}}
\newcommand{\mchisq}{m_{\tilde {\chi}_1^0}^2}
\newcommand{\dtm}{\mbox{${\cos(2\beta) M_Z^2}$}}
\newcommand{\lsp}{\mbox{$\tilde {\chi}_1^0$}}
\newcommand{\Ochi}{\mbox{$\Omega_{\tilde \chi} h^2$}}
\newcommand{\tanb}{\mbox{$\tan \! \beta$}}
\newcommand{\cotb}{\mbox{$\cot \! \beta$}}
\newcommand{\cosb}{\mbox{$\cos \! \beta$}}
\newcommand{\sinb}{\mbox{$\sin \! \beta$}}
\newcommand{\ghcc}{\mbox{$g_{h\tilde{\chi}_1^0 \tilde{\chi}_1^0}$}}
\newcommand{\gHcc}{g_{H\tilde{\chi}_1^0 \tilde{\chi}_1^0}}
\newcommand{\gAcc}{g_{A\tilde{\chi}_1^0 \tilde{\chi}_1^0}}
\newcommand{\gZcc}{\mbox{$g_{Z\tilde{\chi}_1^0 \tilde{\chi}_1^0}$}}
\newcommand{\thb}{\theta_{\tilde b}}
\newcommand{\tht}{\theta_{\tilde t}}
\newcommand{\thq}{\theta_{\tilde q}}
\newcommand{\sto}{\tilde{t}_1}
\newcommand{\sbo}{\tilde{b}_1}
\newcommand{\sqo}{\tilde{q}_1}
\newcommand{\stt}{\tilde{t}_2}
\newcommand{\sbt}{\tilde{b}_2}
\newcommand{\sqt}{\tilde{q}_2}
\renewcommand{\thefootnote}{\fnsymbol{footnote}}

\begin{flushright}
APCTP 97--02 \\
March 1997 \\
\end{flushright}

\vspace*{2.5cm}
\begin{center}
{\Large \bf Recent Developments in Dark Matter Physics}\footnote{Invited
talk presented at the {\it XIIth Symposium of the Department of Atomic
Energy}, Guwahati, Assam, India, Dec.25 1995 to Jan.1 1997} \\
\vspace{5mm}
Manuel Drees \\
\vspace{5mm}
{\it APCTP, 207--43 Cheongryangri--dong, Tongdaemun--gu,
Seoul 130--012, Korea} \\
\end{center}
\vspace{10mm}

\begin{abstract}
After a short review of the arguments for the existence of Particle Dark
Matter in the Universe, I list the most plausible candidates provided by
particle physics, i.e. neutrinos, axions, and WIMPs. In each case I
briefly describe how to estimate the relic density, and discuss attempts at
detecting these particles. At the end I discuss my personal favorite, the
lightest supersymmetric particle, in a little more detail.

\end{abstract}
\clearpage
\setcounter{page}{1}
\pagestyle{plain}
\section*{1) Introduction}
Cosmological Dark Matter (DM) is stuff that at present only manifests itself
through its gravitational interactions; in particular, it does not emit a
detectable amount of electromagnetic radiation at any wavelength. Of course,
the definition of ``detectable'' depends on the sensitivity of the instrument
used for the search, which improves with time. Historically the first evidence
for DM was found as early as 1845, when W.F. Bessel discovered 
irregularities in the proper motion of two stars, Sirius and Procyon \cite{1}.
He concluded that these stars must have ``dark companions'' of roughly their
own mass. These companions were later found to be white dwarves, which at
this relatively short distance are no longer ``dark'' by present standards.

This story of an early, successful DM search is encouraging. However, in
modern understanding cosmological DM refers to stuff that is distributed
over large distance scales, of the order of galactic radii or more
($r \geq 20$ kpc; 1 pc $=3.24$ lyr). This wider distribution makes detection
of such kind of DM much more difficult than finding Bessel's ``dark 
companions''.

Evidence for the existence of galactic DM was found as early as 1922 by
J.H. Jeans, who analyzed the motion of nearby stars transverse to the
galactic plane \cite{1}. He concluded that in our galactic neighborhood
the average density of DM must be roughly equal to that of luminous matter
(stars, gas, dust). Remarkably enough, the most recent estimates, based on
a detailed model of our galaxy, find quite similar results \cite{2}: In
particle physics units, the local DM density must be about
\be \label{e1}
\rho_{\rm DM}^{\rm local} \simeq 0.3 \frac {\rm GeV} {{\rm cm}^3};
\ee
this value is known to within a factor of two or so \cite{2,2a}.

Currently the best evidence for galactic DM comes from the analysis of
galactic rotation curves, i.e. measurements of the velocity with which
things like globular clusters or gas clouds orbit around galaxies. For
a stable circular orbit of radius $r$ from the center of the galaxy,
this velocity is given by
\be \label{e2}
v(r) = \sqrt{ \frac {G_N M(r)} {r} },
\ee
where $G_N$ is Newton's constant and $M(r)$ is the total mass 
inside this orbit. If the mass of the galaxy was concentrated in its
visible part, one would expect $v(r) \propto 1/\sqrt{r}$ at large $r$. Instead,
nearly all of the hundreds of rotation curves that have been studied so
far remain {\em flat} out to the largest observable values of $r$; this
implies $M(r) \propto r$, or $\rho(r) \propto 1/r^2$. One then concludes
that galaxies contain more than ten times more dark than luminous matter.
Mass densities averaged over the entire Universe are usually expressed in units
of the critical or closure density, $\Omega = \rho/\rho_c$ where
$\rho_c \simeq 10^{-29}$ g/cm$^3$; $\Omega = 1$ then corresponds to a flat
Universe. Galactic rotation curves imply
\be \label{e3}
\Omega \geq 0.1; \ \ \ \ ({\rm galactic \ rotation \ curves}).
\ee
This is only a lower bound, since almost all rotation curves remain flat
out to the largest values of $r$ where one can still find objects
orbiting galaxies; we do not know how much further the DM haloes of these
galaxies extend. 

There is considerable evidence for significantly larger $\Omega$
from studies of larger structures, e.g. clusters or superclusters of
galaxies. A fairly conservative observational lower bound on the total mass
density of the Universe is
\be \label{e4}
\Omega > 0.2 \ {\rm to} \ 0.3 \ \ \ \ ({\rm superclusters}).
\ee
Finally, to the best of my knowledge, $\Omega = 1$ is compatible with all
recent observations.\footnote{Determinations of supercluster masses
based on X--ray temperatures give lower values of $\Omega$. However, deriving
the mass density from the measured X--ray energy spectrum is not 
straightforward, and this result seems to contradict more direct determinations
based on gravitational lensing of galaxies lying behind these superclusters.}
This value is favored by ``naturalness'' arguments (since $\Omega = 1$
remains constant in a Friedman--Robertson--Walker cosmology, while
$\Omega \neq 1$ implies an exponential time dependence of $\Omega$,
making its present proximity to 1 difficult to understand), and is also
predicted by most models of cosmological inflation. In contrast, the total
luminous mass density only amounts to
\be \label{e5}
\Omega_{\rm luminous} < 0.01,
\ee
in clear conflict with the bounds (\ref{e3}) and (\ref{e4}).

What could this Dark Matter be? We don't know the answer yet, but we do know 
that not all of it can be ordinary (baryonic) matter. This follows from
analyses of Big Bang nucleosynthesis: Comparing the observed abundances
of ${}^2$H, ${}^3$He, ${}^4$He and ${}^7$Li with predictions, properly
taking into account the chemical evolution of the Universe due to stellar
``burning'', one finds \cite{3}
\be \label{e6}
0.01 \leq \Omega_{\rm baryon} h^2 \leq 0.015.
\ee
Here $h$ is the Hubble constants in units of 100 km/(Mpc$\cdot$sec). A
conservative range for this quantity is $0.4 \leq h \leq 0.9$; most recent
measurements seem to cluster near the lower end of this range, between
0.45 and 0.65 or so \cite{4}. The upper bound in (\ref{e6}) then implies
$\Omega_{\rm baryon} < 0.1$, in mild conflict with the constraint
(\ref{e3}), and in sharp conflict with (\ref{e4}). This, in a nutshell, is
the argument for the existence of exotic (non--baryonic) DM.\footnote{The
upper bound in (\ref{e6}) can be evaded if the baryons are stashed away in
black holes prior to the onset of nucleosynthesis. However, it is not clear
how a large population of such ``primordial'' black holes could have formed.}
Note also that the {\em lower} bound in (\ref{e6}) is in conflict with the
upper bound (\ref{e5}) on the amount of luminous matter, especially if the
recent trend towards a small $h$ holds up. In other words, there is
considerable evidence for baryonic DM as well. The recent discovery of
MACHOs \cite{5} therefore confirms the validity of the overall picture,
including the prediction of non--baryonic DM.

Finally, the important upper bound on the total mass density
\be \label{e7}
\Omega h^2 \leq 1
\ee
follows from the requirement that the Universe must be at least 10 billion
years old, which is a conservative lower bound on the age of the oldest
stars in our galaxy.

\setcounter{footnote}{0}
\section*{2) Candidates for Particle Dark Matter}
As discussed in the previous section, most of the mass of the Universe
seems to be in the form of some exotic, non--baryonic
matter. Fortunately particle physics offers a plethora of candidates
for this dark matter. In this section I will briefly run through this
list, focussing on those candidates whose original {\it raison
d'\`etre} has nothing to do with cosmological considerations.

\subsection*{2a) Light Neutrinos}
Light neutrinos are the only particle DM candidates that are actually known
to exist. An SM neutrino with mass $m_\nu$ contributes \cite{6}
\be \label{e8}
\Omega_\nu h^2 = \frac {m_\nu} {90 \ {\rm eV}}.
\ee
Thus the $\mu$ and/or $\tau$ neutrinos could easily give $\Omega_\nu \sim 1$
without violating laboratory constraints on their masses ($m_{\nu_{\mu}}
\leq 200 \ {\rm keV}, \ m_{\nu_{\tau}} \leq 30$ MeV \cite{7}). This 
appealingly simple solution of the DM problem suffers from two problems,
however. First, the phase space density of neutrinos is
limited by Fermi statistics. This makes it impossible for light neutrinos
to form the dark haloes of dwarf galaxies \cite{8}. Secondly, light 
neutrinos are ``hot'' DM, meaning they were still relativistic when galaxy
formation could have begun (when the causal horizon contained about 1 galactic
mass). Hot DM has a large free--streaming length, which tends to smear out
primordial density perturbations until neutrinos slow down sufficiently.
As a result, models with (mostly) hot DM predict too few old galaxies \cite{9},
{\em if} quantum fluctuations are the ``seed'' of structure formation, as
is assumed in inflationary models. However, in principle dark haloes of
dwarf galaxies could be entirely baryonic without violating the
nucleosynthesis constraint (\ref{e6}); and models where cosmic strings
provide the seed of structure formation can accommodate hot DM \cite{10}.
Finally, it is worth pointing out that some Monte Carlo simulations of
structure formation seeded by quantum fluctuations indicate \cite{11} that the
observed hierarchy of structures is reproduced best by a mix with
$\Omega_{{\rm hot \ DM}} \simeq 0.25, \ \Omega_{{\rm cold \ DM}} \simeq 0.7$,
and $\Omega_{\rm baryon} \simeq 0.05$.

Light neutrino DM also has the ``practical'' disadvantage that it is almost
impossible to detect. By now these neutrinos are nonrelativistic, so their
annihilation or scattering can only release a few (tens of) eV of energy.
To my knowledge no scheme for their detection has yet been proposed.
It is also almost inconceivable that the range of masses indicated by 
eq.(\ref{e8}) can be probed directly (kinematically) for the $\mu$ and
$\tau$ neutrino. Our best hope is that massive neutrinos might mix with
each other, leading to in principle observable neutrino flavor oscillations,
which could allow us to determine the differences of their squared masses.

\subsection*{2b) WIMPs}
Weakly interacting massive particles (WIMPs) are particles with masses
roughly between 10 GeV and a few TeV, and with cross sections of approximately
weak strength. The reason for considering such particles as DM candidates
rests on a curious ``coincidence'': Their present relic density is 
approximately given by \cite{2a}
\be \label{e9}
\Omega_{\rm WIMP} h^2 \simeq \frac { 0.1 \ {\rm pb} \cdot c} {\langle 
\sigma_A v \rangle }.
\ee
Here $c$ is the speed of light, $\sigma_A$ is the total annihilation cross
section of a pair of WIMPs into SM particles, $v$ is the relative velocity
between the two WIMPs in their cms system, and $\langle \dots \rangle$
denotes thermal averaging. This follows from the
fact that WIMPs are non--relativistic already when they drop out of 
equilibrium with the hot thermal ``soup'' of SM particles (``freeze--out''),
which occurs at temperature $T_F \simeq m_{\rm WIMP}/20$ almost 
independently of the properties of the WIMP. One can then derive eq.(\ref{e9})
by requiring that the WIMP annihilation rate $\Gamma = n_{\rm WIMP}
\langle \sigma_A v \rangle$ be equal to the expansion rate of the Universe
$H$ at $T = T_F \simeq m_{\rm WIMP}/20$.\footnote{In a more complete
treatment the (logarithmic) dependence of $T_F$ on $\sigma_A$ has to be
included, which leads to a set of coupled equations that can easily be
solved by iteration \cite{2a}.} Notice that the constant in eq.(\ref{e9}),
0.1 pb, contains factors of the Planck mass, the current temperature of
the microwave background, etc; it is therefore quite intriguing that it
``happens'' to come out near the typical size of weak interaction cross
sections. 

Since WIMPs annihilate with very roughly weak interaction strength, it
is natural to assume that their interaction with normal matter is also
approximately of this strength. This raises the hope of detecting
relic WIMPs directly \cite{2a}, by observing their scattering
off nuclei in a detector. The energy deposition can be several (tens
of) keV. This is quite easily detected; however, care has to be taken
to suppress backgrounds from ambient or intrinsic radioactivity, and
from cosmic rays. There are now about 10 different groups searching
for such signals \cite{12}.

Alternatively one can look for signals for ongoing WIMP annihilation.
The WIMP density in free space is too small to give a detectable
signal (except possibly near the center of our galaxy
\cite{2a}). However, the same scattering processes that might allow to
detect WIMPs directly can also lead to WIMP capture by celestial
bodies, in particular the Earth or Sun. This happens if a WIMP loses
so much energy in a scattering reaction that it becomes
gravitationally bound. Eventually WIMPs will then become sufficiently
concentrated near the center of these bodies to annihilate with
significant rate. Once equilibrium is reached, the annihilation rate
will simply be half the capture rate (half, since each annihilation
event destroys two WIMPs). Since these annihilations occur near the
center of the Earth or Sun, the only possibly detectable annihilation
products are neutrinos, in particular muon (anti)neutrinos. The signal
\cite{2a} is therefore muons pointing back towards the center of the
Earth or Sun; the energy spectrum of these muons is also expected to
be different from that produced by atmospheric (cosmic ray induced)
neutrinos.

The perhaps most obvious WIMP candidate is a heavy neutrino. However,
an $SU(2)$ doublet neutrino will have too small a relic density if its
mass exceeds a few GeV, as required by LEP data. One can suppress the
annihilation cross section, and hence increase the relic density,
by postulating mixing between a heavy $SU(2)$ doublet and some ``sterile''
$SU(2) \times U(1)_Y$ singlet neutrino. However, one also has to require the
neutrino to be stable; it is not obvious why a massive neutrino should not
be allowed to decay.

In supersymmetric models with exact R--parity the lightest
supersymmetric particle (LSP) is absolutely stable. Searches for
exotic isotopes \cite{7} then imply that it has to be neutral. This
leaves basically two candidates in the ``visible sector'', a sneutrino
and a neutralino. Sneutrinos again have quite large annihilation cross
sections; their masses would have to exceed several hundred GeV for
them to make good DM candidates. This is uncomfortably heavy for the
lightest sparticle, in view of naturalness arguments.\footnote{In the
recently popular models with gauge--mediated SUSY breaking the
lightest ``messenger sneutrino'' could well be stable, and
sufficiently massive, even though it is not the LSP \cite{13}; note
that most of its mass comes from supersymmetric terms, i.e. does not
contribute to SUSY breaking.} Further, the negative outcome of various
WIMP searches rules out sneutrinos as primary component of the DM halo
of our galaxy \cite{14}. In contrast, the lightest neutralino still
can make a good DM candidate; this will be discussed in a little more
detail in Sec.~3.

\subsection*{2c) Axions}
Axions \cite{15} are hypothetical pseudo--Goldstone bosons of a 
spontaneously broken new ``Peccei--Quinn'' (PQ) symmetry that allows to
``rotate away'' the CP--violating $\theta$ parameter of QCD; in other
words, axions have been introduced to solve the strong CP problem.
They are not completely massless, since the PQ symmetry is not only broken
by the vev of the scalar partner of the (pseudoscalar) axion, but also
``explicitly'' by nonperturbative QCD effects. As a result,
\be \label{e10}
m_a \simeq 0.6 \ {\rm meV} \cdot \frac {10^{10} \ {\rm GeV}} {f_a},
\ee
where $f_a$ is the scale of PQ symmetry breaking and $m_a$ is the mass of
the axion. Relic axions are produced athermally during the QCD phase
transition. If this is the main source of relic axions, then \cite{16}
\be \label{e11}
\Omega_a h^2 \simeq 0.9 \left( \frac {f_a} {10^{12} \ {\rm GeV}}
\right)^{1.18} \cdot \overline{ \theta^2_i},
\ee
where $\overline{\theta_i^2}$ is the average initial value of the axion
field (written as a phase). This quantity is ``naturally'' of order unity,
but it might be ``accidentally'' much smaller; in this case $f_a$ would
need to be correspondingly larger for axions to form significant amounts
of DM. On the other hand, axion models often predict the existence of
cosmic strings. In such scenarios the emission of axions from these strings
is the main source of relic axions, and one would need a smaller $f_a$
in order not to violate the bound (\ref{e7}). Finally, a series of
laboratory and astrophysical constraints implies
\be \label{e12}
f_a \geq 5 \cdot 10^9 \ {\rm GeV}.
\ee

In spite of their small mass, axions are cold DM, since they were
produced athermally. They are much too light for the techniques used
in WIMP searches to be applicable here. Instead, one looks for $a
\rightarrow \gamma$ conversion in a strong magnetic field. Such a conversion
proceeds through the loop--induced $a \gamma \gamma$ coupling, whose
strength $g_{a \gamma \gamma}$ is an important parameter of axion models.
Currently two axion search experiments are taking data. They both employ
high quality cavities, since the cavity ``q--factor'' enhances the
conversion rate on resonance, i.e. if $m_a c^2 = h \nu_{\rm res}$. One
then needs to scan the resonance frequency in order to cover a significant
range in $m_a$ or, equivalently, $f_a$. The bigger of the two experiments,
situated at the LLNL in California, started taking data in the first half
of 1996. The analysis of the first data set, covering about 1/3 of a
decade in $m_a$ values, should be published soon \cite{17}.

The LLNL experiment uses ``conventional'' electronic amplifiers to
enhance the conversion signal, albeit very sophisticated ones with
exceedingly low noise temperature. In contrast, a smaller experiment now
under way in Kyoto, Japan \cite{18} uses Rydberg atoms (atoms
excited to a very high state, $n \simeq 230$) to detect the microwave
photons that would result from axion conversion. In spite of the
significantly smaller volume of this experiment, this allows them to
reach better sensitivity than the LLNL experiment. While the latter
can only probe models where $g_{a \gamma \gamma}$ is near the upper end
of the expected range, the former will test all axion models that have been
proposed to far. However, the tuning range, i.e. the range of $m_a$ values
that can be covered, is much smaller for the Kyoto experiment. This
experiment started taking data at the very end of 1996, and should publish
its results this year.

\subsection*{2d) Other Candidates}
There are DM candidates which do not belong to any of the classes
discussed so far. One example are new ``baryons'', which interact
strongly with each other, but with gauge group different from the
standard $SU(3)_c$. In analogy with the ordinary strong interactions,
one usually assumes that these new ``baryons'' can annihilate into
somewhat lighter, unstable new ``mesons'', with a cross section that
more or less saturates unitarity limits \cite{19}. Such ``baryons''
will have relic density $\Omega \sim 1$ if their mass is ${\cal
O}(500)$ TeV. ``Baryons'' with mass in this range could, e.g., exist
\cite{13} in the ``hidden sector'' of models with gauge--mediated SUSY
breaking. Since these particles are singlets under the SM gauge group,
their (loop--induced) interactions with ordinary matter are
exceedingly feeble, making them almost impossible to detect
experimentally. Similar candidates can also exist in extended
technicolor models; however, these techni--baryons are more easily
detected by WIMP search experiments \cite{20}.

Another possible DM candidate is the gravitino, the spin--3/2
superpartner of the graviton. This will usually be the LSP if the SUSY
breaking scale\footnote{This scale is given by the expectation value
of the largest SUSY--breaking $F$ or $D$ term; it should not be
confused with the mass scale of ordinary sparticles, which for
phenomenological reasons has to lie in the (few) hundred GeV to TeV
range.} is significantly below $\sqrt{M_Z M_{Pl}} \sim 10^{10}$
GeV. Since gravitinos only interact gravitationally, they are still
relativistic at freeze--out. However, the interaction strength of
``longitudinal'' ($S_z = \pm 1/2$) gravitinos scales inversely
proportional to the gravitino mass $m_{\widetilde G}$. As a result,
the relic density $\Omega_{\widetilde G} \propto m_{\widetilde G}$,
and becomes ${\cal O}(1)$ for $m_{\widetilde G} \simeq 0.85$ keV
\cite{21}. Even though gravitinos are relativistic at freeze--out, by
the time structure formation starts they have slowed down sufficiently
to form ``warm'' DM, which resembles cold DM for most
purposes. Finally, one could produce an additional ``hot'' (really,
athermal) gravitino component from neutralino decays. While generating
both (almost) hot and (almost) cold DM from the same particle looks
like a neat trick, one needs slepton masses well in excess of 1 TeV
for the ``hot'' component to be significant \cite{21}, which makes
this scheme rather unattractive. Relic gravitinos are probably the
most difficult to detect of all DM candidates.

\setcounter{footnote}{0}
\section*{3) The Lightest Neutralino}
Let me now discuss my personal favorite DM candidate, the lightest
neutralino, in a little more detail. It is my favorite since Supersymmetry
is the in my opinion best motivated extension of the SM. When coupled with
Grand Unification, SUSY models usually predict the lightest neutralino to
be the lightest sparticle of the visible sector. While the possibility of
having an even lighter ``hidden sector'' sparticle (like the gravitino)
cannot a priori be excluded, there is no good reason for such sparticles
to be light enough to satisfy the bound (\ref{e7}) (e.g., $m_{\wt G} 
< 1$ keV; see Sec.~2c), given that the SUSY breaking scale is
associated with the weak scale. Finally, while not easy to detect, there
is hope that relic neutralinos can eventually be proven unambiguously
(not) to exist. 

The Minimal Supersymmetric Standard Model (MSSM) \cite{22}, to which I will
restrict myself here, contains four neutralino current states: The
superpartners of the $U(1)_Y$ gauge boson (the ``bino'' $\wt{B}$), of the
neutral $SU(2)$ gauge boson (neutral ``wino'' $\wt{W}_3$), and of the two
neutral Higgs bosons needed in any SUSY model \cite{22} (``higgsinos''
$\wt{h}_1^0$ and $\wt{h}_2^0$, with $Y_{H_1} = - Y_{H_2} = -1/2$). Since the
electroweak gauge symmetry is broken, these current states mix to form 
four Majorana mass eigenstates. At tree--level the mass matrix in the
basis $(\wt{B}, \wt{W}_3, \wt{h}_1^0, \wt{h}_2^0)$ is given by
\be \label{e13}
{\cal M}_0 = \mbox{$ \left( \begin{array}{cccc}
M_1 & 0 & - M_Z \cos \! \beta \sin \! \theta_W & M_Z \sin \! \beta \sin \!
\theta_W \\
0 & M_2 &   M_Z \cos \! \beta \cos \! \theta_W & -M_Z \sin \! \beta \cos \!
\theta_W \\
- M_Z \cos \! \beta \sin \! \theta_W & M_Z \cos \! \beta \cos \! \theta_W & 
0 & -\mu  \\
M_Z \sin \! \beta \sin \! \theta_W & -M_Z \sin \! \beta \cos \! \theta_W &
-\mu  & 0
\end{array} \right) $} 
\ee
Here, $M_1$ and $M_2$ are SUSY breaking gaugino masses, $\mu$ is a 
supersymmetric Higgs(ino) mass parameter, and $\tanb = \langle H_2^0
\rangle / \langle H_1^0 \rangle$ is the ratio of vevs. The number of 
free parameters is reduced if one assumes that the gaugino masses unify, just
like the MSSM gauge couplings seem to do \cite{23}; this implies
\cite{22} for the running masses:
\be \label{e14}
M_1 = \frac {5}{3} \tan^2 \theta_W M_2 \simeq 0.5 M_2,
\ee
where the second equality holds near the weak scale.

The size of the entries of the neutralino mass matrix (\ref{e13}) that mix
gaugino and higgsino states is bounded by the mass of the $Z$ boson. 
This is not surprising, since such mixing can only occur once $SU(2)
\times U(1)_Y$ is broken, and $M_Z$ characterizes the strength of gauge
symmetry breaking in the MSSM just as it does in the SM. On the other hand,
the size of the diagonal entries $M_1, \ M_2$ and $\mu$ is not (yet)
known. It is useful to study two limiting situations, where $|\mu|$ is
either significantly larger or significantly smaller than the gaugino
masses $M_{1,2}$.

In the first case, $|\mu| > M_2$, the lightest neutralino \lsp\ is
mostly a gaugino. If the unification relation (\ref{e14}) holds, \lsp\
will be mostly a photino if $M_2^2 \ll M_Z^2$, turning into a bino
if $M_1 > M_Z$. The $Z - \lsp - \lsp$ coupling is then proportional to the
square of the small higgsino component of \lsp, while the Higgs--\lsp--\lsp\
couplings are linear in this small component. However, the $\lsp - f -
\tilde{f}$ couplings have full [$U(1)_Y$ or $U(1)_{\rm em}$] gauge
strength. Unless $\mchi \simeq M_Z/2$ or $\mchi \simeq m_{\rm Higgs}/2$,
in which case $s-$channel diagrams are ``accidentally'' enhanced, 
annihilation of these gaugino--like LSPs therefore proceeds dominantly
through sfermion exchange in the $t-$ or $u-$channel, leading to $f 
\bar{f}$ final states. Since the cross section is proportional to
the fourth power of the (hyper)charge, the dominant contribution comes
from the exchange of (right--handed) charged sleptons. It has been known
for quite some time \cite{24} that this leads to relic density $\Ochi
\sim 1$ for very reasonable SUSY parameters. Specifically, for a bino--like
LSP away from $s-$channel poles, one finds \cite{25}
\be \label{e15}
\Ochi \simeq \frac {\Sigma^2} { (1 \ {\rm TeV})^2 \mchisq } \cdot
\frac {1} { \left( 1 - \mchisq/\Sigma \right)^2 +
m_{\tilde {\chi}_1^0}^4/\Sigma^2 },
\ee
where $\Sigma = \mchisq + m^2_{\tilde{l}_R}$, and I have assumed three
degenerate $SU(2)$ singlet sleptons $\tilde{l}_R$. As advertised, this
gives a cosmologically interesting relic density for sparticle masses in
the (few) hundred GeV range.

This scenario is favored in models with GUT boundary conditions, if
one assumes degeneracy of all soft breaking scalar masses at this very
high energy scale. The electroweak gauge symmetry is then broken
radiatively, and $|\mu|$ usually comes out quite large \cite{26}, due
to the large top mass. On the other hand, since the LSP couples only
weakly to $Z$ and Higgs bosons, its scattering cross section off
ordinary matter is quite small. This illustrates that the ``natural''
assumption of a weak--scale scattering cross section, given a
weak--scale annihilation cross section, can be off by a large factor.
We saw that annihilation in this example proceeds through slepton
exchange. However, slepton exchange can only contribute to LSP
scattering off electrons, the cross section for which is suppressed by
a factor $(m_e/m_N)^2 < 10^{-6}$ compared to the one for scattering
off nuclei.  This latter process can proceed through squark exchange,
but this entails a suppression factor $\left( Y_{\tilde q}
m_{\tilde{e}_R} / Y_{\tilde{e}_R} m_{\tilde q} \right)^4$, which can
be as small as $10^{-4}$ in many models. The dominant contribution to
LSP scattering off nuclei therefore usually comes from scalar Higgs
exchange, if the LSP is gaugino--like \cite{2a}. Since the relevant
coupling is quite small, as discussed earlier, such relic LSPs are
quite difficult to detect. For example, the direct detection rate in a
Germanium detector is typically \cite{27} $10^{-4}$ to $10^{-2}$
evts/(kg$\cdot$day) for $\mu < 0$, and about five times larger for
$\mu > 0$. This has to be compared with a current sensitivity which
does not extend below $10^{+1}$ evts/(kg$\cdot$day); event
next--generation experiments only aim for sensitivity around $10^{-1}$
evts/(kg$\cdot$day). However, reaching the necessary sensitivity is
not inconceivable. Indirect detection (through LSP annihilation in the
Earth or Sun) is also very challenging in this scenario.

In the opposite limit, $M_1^2, \ M_2^2 \gg \mu^2$, the lightest neutralino is
dominantly a higgsino. Since the higgsino mass term $\mu$ in eq.(\ref{e13})
connects $\wt{h}_1^0$ and $\wt{h}_2^0$, \lsp\ is a combination of both
current eigenstates:
\be \label{e16}
\lsp \simeq \frac {1}{\sqrt{2}} \left( \wt{h}_1^0 - {\rm sign}(\mu)
\wt{h}_2^0 \right) \ \ \ \ \ \ \ \ (M_1^2, \ M_2^2 \rightarrow \infty).
\ee
Since the Higgs$-\lsp-\lsp$ couplings probe both the higgsino and the
gaugino components of \lsp, these couplings again vanish in the limit
(\ref{e16}). For finite $M_1, \ M_2$, \lsp\ is not exactly given by 
eq.(\ref{e16}); there are corrections of order $M_W^2/(\mu M_2)$, hence
the tree--level Higgs$-\lsp-\lsp$ couplings will be of this order. 
Furthermore, the $Z-\lsp-\lsp$ coupling also vanishes in the limit
(\ref{e16}), since it is \cite{22} proportional to the
{\em difference} of the squares of the two higgsino components.
Finally, the $\lsp-f-\tilde{f}$ couplings are now Yukawa couplings,
and hence quite small except for $f=t$ (and $f=b$ or $\tau$, if
$\tanb \gg 1$). As a result, a higgsino--like LSP with mass $\mchi < M_W$
has a very small annihilation cross section, unless $\tanb \gg 1$.

One might therefore think that such a state has a very large relic density,
see eq.(\ref{e9}). This is, however, not correct. The reason is that the
MSSM contains three higgsino--like states if $M_2^2 \gg \mu^2$: The
second lightest neutralino is a higgsino state orthogonal to the LSP
(\ref{e16}), and the lightest chargino is also higgsino--like. All three
states have mass $|\mu|$, up to corrections of order $M_W^2/M_2$. For large
$M_2$ the mass splitting between these three states is therefore much
smaller than the splitting between \lsp\ and the (nearly) massless states of
the SM. As a result, the three higgsino--like states remain in {\em relative}
thermal equilibrium well after the entire SUSY sector has frozen out of
equilibrium with the SM states. The reason is that reactions of the type
$f \tilde {\chi}_i^0 \leftrightarrow f \tilde {\chi}_j^0$ and
$f \tilde {\chi}_i^0 \leftrightarrow f' \tilde {\chi}_1^\pm$ occur much
more frequently than reactions like $\tilde{\chi}_i^0 \tilde{\chi}_j^0
\leftrightarrow f \bar{f}$ if $\left| m_{\tilde{\chi}_i} -
m_{\tilde{\chi}_j} \right| \ll \left| m_{\tilde{\chi}_i} - m_f \right|$.

Under such circumstances co--annihilation of the LSP with one of the
heavier higgsinos becomes important \cite{28}. That is, one also has to
consider reactions like $\tilde{\chi}_1^0 \tilde{\chi}_2^0 \rightarrow
f \bar{f}$ and $\tilde{\chi}_1^0 \tilde{\chi}_1^\pm \rightarrow f \bar{f'}$
when estimating the relic density. Notice that the $Z-\lsp-\tilde{\chi}_2^0$
and $W^\pm-\lsp-\tilde{\chi}_1^\mp$ couplings have full gauge strength in
this scenario; the co--annihilation reactions therefore have quite large
cross sections. As a result, the relic density of higgsino--like LSPs is
actually quite small \cite{29}.

So far the analysis was based on tree--level results. Technically, the
degeneracy of the three higgsino states in the limit $M_2 \rightarrow \infty$
hinges on the fact that the $\wt{h}_1^0 \wt{h}_1^0$ and $\wt{h}_2^0
\wt{h}_2^0$ entries of the mass matrix (\ref{e13}) vanish; this is a
consequence of $SU(2) \times U(1)_Y$ gauge invariance. Since this gauge
invariance is broken, we might expect such entries to be generated at the
one--loop level. This is indeed the case \cite{30}, with the dominant
contribution coming from heavy quark -- squark loops \cite{31}. In
particular, $t-\tilde{t}$ loops generate an $\wt{h}_2^0 \wt{h}_2^0$ entry
of order \cite{31,32}
\be \label{e17}
\delta_{44} \simeq \frac {3 G_F} {8 \pi^2} \frac {m_t^3} {\sin^2 \beta}
\sin(2 \theta_{\tilde t}) \log \frac {m_{\tilde{t}_2}} {m_{\tilde{t}_1}},
\ee
where $G_F$ is the Fermi constant, $\tilde{t}_1$ and $\tilde{t}_2$ are
the two stop mass eigenstates, and $\theta_{\tilde t}$ is the 
$\tilde{t}_L - \tilde{t}_R$ mixing angle. Notice that the correction
$\delta_{44}$ vanishes if the two stop eigenstates are unmixed or degenerate
in mass. Numerically, $\delta_{44}$ can be as large as $\sim 8$ GeV
\cite{32}. Note that $m_{\tilde{\chi}_2^0} - \mchi \simeq \delta_{44}$
and $m_{\tilde{\chi}_1^\pm} - \mchi \simeq \delta_{44}/2$ in the limit
$M_2 \rightarrow \infty$. These mass splittings appear in the expression for
the relic density due to co--annihilation in the form of exponential
Boltzmann factors, $\exp \left(- \Delta m_{\tilde \chi}/T_F \right)
\simeq \exp \left( - 20 \Delta m_{\tilde \chi} / \mchi \right)$. As a result,
the loop corrections to the mass splittings can change the estimate of
the LSP relic density by up to a factor of five in either direction, if
\lsp\ is a nearly pure higgsino state \cite{32}. If the sign of the
correction is such that it increases the mass splittings, a state with
99.9\% higgsino purity can form galactic DM ($\Ochi \geq 0.025$), while
a state with 99.5\% higgsino purity can form all cold DM in models with
mixed cold and hot DM ($\Ochi \geq 0.15$).

Closely related $t-\tilde{t}$ loop corrections can also have
significant impact on the coupling of higgsino--like LSPs to $Z$ and
Higgs bosons, and thus on the expected LSP detection rate \cite{32}. In
particular, for $\mu < 0$ the coupling to the lighter scalar Higgs
boson can increase tenfold, which increases the estimate of the LSP
detection rate by fully two orders of magnitude if this rate is
dominated by scattering off spinless nuclei (e.g., direct detection
using heavy nuclei, or capture in the Earth). The reason is that for this
sign of $\mu$ the tree--level coupling is not only suppressed by the small
size of higgsino--gaugino mixing for $M_2^2 \gg \mu^2$, it also suffers
additional ``accidental'' cancellations. Since the tree--level coupling is
so small, loop corrections can even reverse its sign. As a result, the
loop--corrected coupling, and the LSP scattering cross section off spinless
nuclei, might vanish completely.\footnote{The zero of the scattering matrix
element occurs for slightly different parameter combinations than the zero
of the $\lsp-\lsp-h^0$ coupling, since the matrix element also gets 
contributions from heavy Higgs and squark exchange \cite{2a}.} Fortunately
the cross section for scattering off nuclei with non--vanishing spin remains
finite in this case; it is, however, very small. Even for maximal positive
correction, the LSP detection rate in a Germanium detector remains below
$10^{-2}$ evts/(kg$\cdot$day) for $\mu<0$ and higgsino--like LSP; this is
not much better than for bino--like LSP. However, for $\mu>0$ viable
solutions can be found where the detection rate exceeds 0.1 
evts/(kg$\cdot$day); this necessitates a quite substantial gaugino component
of the LSP (several percent at least).

Higgsino--like states with mass exceeding $M_W$ again have quite small relic
density, since they have large annihilation cross sections into $W^+ W^-$
and $ZZ$ final states. Even in the absence of co--annihilation one needs
$\mchi > 250$ (600) GeV for such an LSP to form galactic (all cold) DM.
In this case the total co--annihilation cross sections are not much larger
than the $\lsp\lsp$ annihilation cross section. Nevertheless co--annihilation
will increase these lower bounds, possibly by as much as a factor of two.

Finally, if the unification condition (\ref{e14}) does not hold, the LSP
might also be $\wt{W}_3$--like. This again leads to a tiny relic density
\cite{33}. The culprit is again co--annihilation, this time exclusively
with the nearly degenerate lighter chargino. The tree--level mass splitting
in this case is even smaller than that between the higgsino--like states,
and loop corrections only amount to less than 100 MeV here. Furthermore,
the $W^\pm - \lsp - \tilde{\chi}^\mp$ coupling is now a triplet coupling,
rather than a doublet coupling as in case of higgsino--like LSP. As a result,
the relic density \Ochi\ is below $10^{-4}$ for $\mchi \leq M_W$. A
neutral wino--like LSP is therefore definitely not a good DM candidate.

\section*{4) Summary and Conclusions}
There are compelling arguments for the existence of exotic dark matter in
the Universe. We don't presently know just what this DM is made of, but
there are many particle physics candidates standing in line to fill
this vacancy. Out of these the best motivated ones are light neutrinos,
axions, and neutralinos.

Light neutrinos are known to exist, but it is not clear whether they have
the required mass, in the range of a few (tens of) eV. Unfortunately testing
whether relic neutrinos form all or part of the required DM is exceedingly
difficult, due to the minuscule size of the relevant cross sections and the
small amount of the energy that could possibly be deposited by them. In case
of $\mu$ and $\tau$ neutrinos a direct kinematical measurement of eV scale
masses using lab experiments is also essentially hopeless. Such experiments
might teach us something about differences of squares of neutrino masses,
{\em if} the different neutrino flavor eigenstates mix sufficiently
strongly. However, the only reasonably direct way of measuring eV scale
$\nu_\mu$ and $\nu_\tau$ masses that I can think of is through precise
timing of neutrino pulses emitted by supernovae; unfortunately we may have
to wait decades for the next sufficiently close explosion. Finally, recall
that current conventional wisdom disfavors neutrinos as dominant DM component.

Axions have been postulated in order to solve the strong CP problem. In my
view they suffer from the theoretical, or rather aesthetical, problem that
this explanation does not constrain the scale $f_a$ at all; in principle
it could be anywhere between $\Lambda_{\rm QCD}$ and $M_{Pl}$. 
Laboratory searches and astrophysical constraints exclude the lower half
of this range (on a logarithmic scale), while the bound (\ref{e7}) on
the relic density excludes, or at least strongly disfavors, very large values
of $f_a$. At present one or two decades in between are still allowed.
This can be interpreted in two ways. If you don't like axions, you might
argue that they suffer a finetuning problem, since most of the a priori
allowed range is already excluded. If you do like axions, you can emphasize
the fact that in this allowed window, axions probably form at least a
significant fraction of all DM. In any case, axions are quite unique in
particle physics in that relic axions are the {\em only} axions that we
can possibly detect. Indeed, we'll probably know within a decade or so
whether axions form a significant fraction of the dark halo of our own
galaxy.

However, other solutions of the strong CP problem have been
suggested. In fact, some people \cite{34} (including myself) consider
this problem to ``only'' be one aspect of the overall flavor problem,
so introducing a special particle just for this one facet of the
problem seems rather extravagant. In contrast, Supersymmetry is the so
far only solution of the hierarchy problem that is at least potentially
fully realistic (in agreement with all existing data). Moreover, in the
simplest viable models the lightest neutralino emerges almost automatically
as an attractive DM candidate. The only assumption one has to make is that
of minimality -- that is, that certain couplings, which seem entirely
unnecessary, are indeed absent in the Lagrangian, so that R--parity is
conserved. 

We saw in Sec.~3 that in general the lightest neutralino can come in
different forms. A photino-- or bino--like state makes the most
natural DM candidate in the sense that it has a cosmologically
interesting relic density for a fairly wide region of parameter
space. A smaller window exists also for a light higgsino--like LSP,
partly due to radiative corrections which can be quite important in
this case. Such relic neutralinos are probably quite difficult to
detect, although the task is at least not as hopeless as for certain
other DM candidates discussed in Sec.~2d. Fortunately sparticles
should leave plenty of tell--tale signatures in collider
experiments. In particular, the forthcoming LHC at CERN should be able
to unambiguously test the idea of weak--scale Supersymmetry
\cite{35}. However, even if SUSY is first discovered at colliders,
relic LSP searches do not become less important.  For one thing, their
mere detection would immediately raise the lower bound on their
lifetime from something like $10^{-7}$ seconds, which is an optimistic
guess for the sensitivity of collider experiments, to about $10^{+18}$
years, an improvement of some 32 orders of magnitude! Even more
exciting, detecting relic neutralinos, or any other kind of exotic
dark matter, would finally tell us what gives the Universe (most of)
its mass.


\begin{thebibliography}{99}
 
\bibitem{1}
For a brief but delightful history of DM, see V. Trimble, in Proceedings of
the {\it First International Symposium on Sources of Dark Matter in the
Universe}, Bel Air, California, 1994; published by World Scientific,
Singapore (ed. D.B. Cline).

\bibitem{2}
M.S. Turner, E.I. Gates and G. Gyuk, astro--ph 9601168.

\bibitem{2a}
For a review, see G. Jungman, M. Kamionkowski and K. Griest, Phys. Rep.
{\bf 267}, 195 (1996).

\bibitem{3}
For a review, see S. Sarkar, Rep. Prog. Phys. {\bf 59}, 1493 (1996).

\bibitem{4}
D.N. Schramm, talk presented at the {\it RESCEU Symposium on Dark Matter in
the Universe and its Direct Detection}, Tokyo, November 1996; to appear
in the proceedings (ed. M. Minowa).

\bibitem{5}
MACHO collab., C. Alcock et al., Nature {\bf 365}, 621 (1993), and
W. Sutherland et al., astro--ph 9611059;
EROS collab., E. Aubourg et al., Nature {\bf 365}, 623 (1993).

\bibitem{6}
R. Cowsik and J. McClelland, Phys. Rev. Lett. {\bf 29}, 669 (1972).

\bibitem{7}
Particle Data Group, R.M. Barnett et al., Phys. Rev. {\bf D54}, 1 (1996).

\bibitem{8}
S. Tremaine and J.E. Gunn, Phys. Rev. Lett. {\bf 42}, 407 (1979).

\bibitem{9}
S.D.M. White, C.S. Frenk and M. Davis, Ap. J. {\bf 274}, L1 (1983).

\bibitem{10}
V. Zanchin, J.A.S. Lima and R. Brandenberger, Phys. Rev. {\bf D54},
7129 (1996).

\bibitem{11}
E.g., J.R. Primack and J.A. Holtzmann, Ap. J. {\bf 405}, 428 (1993).

\bibitem{12}
An up--to--date overview of this field can be found in the Proceedings
of the RESCEU Dark Matter conference \cite{4}.

\bibitem{13}
S. Dimopoulos, G.F. Giudice and A. Pomarol, hep--ph 9607225, Phys. Lett.
{\bf B389}, 37 (1996).

\bibitem{14}
The currently best published WIMP search limits are: \\
For direct detection: J.J. Quenby et al., Phys. Lett. {\bf B351}, 70 (1995);
\\ For indirect detection: Baksan collab., as presented at TAUP95,
Valencia, Spain, September 1995.

\bibitem{15}
For a review, see M.S. Turner, Phys. Rep. {\bf 197}, 67 (1990).

\bibitem{16}
R.A. Battye and E.P.S. Shellard, as in \cite{1}.

\bibitem{17}
L. Rosenberg, as in \cite{4}.

\bibitem{18}
I. Ogawa, as in \cite{4}.

\bibitem{19}
K. Griest and M. Kamionkowski, Phys. Rev. Lett. {\bf 64}, 615 (1990).

\bibitem{20}
J. Bagnasco, M. Dine and S. Thomas, hep--ph 9310290, Phys. Lett. {\bf B320},
99 (1994).

\bibitem{21}
S. Borgani, A. Masiero and M. Yamaguchi, hep--ph 9605222, Phys. Lett.
{\bf B386}, 189 (1996).

\bibitem{22}
For reviews, see H.P. Nilles, Phys. Rep. {\bf 110}, 1 (1984);
H.E. Haber and G.L. Kane, Phys. Rep. {\bf 117}, 75 (1985).

\bibitem{23}
U. Amaldi, W. deBoer and H. F\"urstenau, Phys. Lett. {\bf B260},
447 (1991);
P. Langacker and M. Luo, Phys. Rev. {\bf D44}, 817 (1991);
J. Ellis, S. Kelley and D.V. Nanopoulos, Phys. Lett. {\bf B260},
131 (1991).

\bibitem{24}
H. Goldberg, Phys. Rev. Lett. {\bf 50}, 1419 (1983);
J. Ellis, J. Hagelin, D.V. Nanopoulos, K. Olive and M. Srednicki, Nucl. Phys.
{\bf B238}, 453 (1984);
L. Roszkowski, Phys. Lett. {\bf B262}, 59 (1991), and {\bf B278}, 147 (1992).

\bibitem{25}
M. Drees and M.M. Nojiri, Phys. Rev. {\bf D47}, 376 (1993).

\bibitem{26}
For a review, see M. Drees and S.P. Martin, hep--ph 9504324.

\bibitem{27}
M. Drees and M.M. Nojiri, Phys. Rev. {\bf D48}, 3483 (1993);
P. Nath and R. Arnowitt, hep--ph 9701301.

\bibitem{28}
K. Griest and D. Seckel, Phys. Rev. {\bf D43}, 3191 (1991).

\bibitem{29}
S. Mizuta and M. Yamaguchi, Phys. Lett. {\bf B298}, 120 (1993).

\bibitem{30}
D. Pierce and A. Papadopoulos, Phys. Rev. {\bf D50}, 565 (1994), and
Nucl. Phys. {\bf B430}, 278 (1994);
A.B. Lahanas, K. Tamvakis and N.D. Tracas, Phys. Lett. {\bf B324}, 387 (1994).

\bibitem{31}
G.F. Giudice and A. Pomarol, Phys. Lett. {\bf B372}, 253 (1996).

\bibitem{32}
M. Drees, M.M. Nojiri, D.P. Roy and Y. Yamada, hep--ph 9701219.

\bibitem{33}
S. Mizuta, D.Ng and M. Yamaguchi, Phys. Lett. {\bf B300}, 96 (1993);
C.--H. Chen, M. Drees and J.F. Gunion, Phys. Rev. {\bf D55}, 330 (1997).

\bibitem{34}
See e.g. S.M. Barr, hep--ph 9612396, and references therein.

\bibitem{35}
See e.g. H. Baer et al., hep--ph 9503479, and references therein.

\end{thebibliography}
\end{document}